\documentclass[a4paper,amsfonts, amssymb, amsmath, reprint,onecolumn, showkeys, nofootinbib,notitlepage]{revtex4-1}
\usepackage[english]{babel}
\usepackage[utf8]{inputenc}
\usepackage{graphicx}
\usepackage{dcolumn}
\usepackage{bm}
\usepackage{hyperref} 
\begin{document}

\title{Status of kinematic cosmology with SN Ia: \\ JLA, Pantheon and future constraints with LSST}

\author{Caroline Heneka}
 \email{caroline.heneka@sns.it}
\affiliation{Scuola Normale Superiore, Piazza dei Cavalieri 7, 56126 Pisa, Italy \\
 Institut f\"ur Theoretische Physik, Ruprecht-Karls-Universit\"at Heidelberg, Philosophenweg 16, 69120 Heidelberg, Germany
}%

\date{\today}

\begin{abstract}
In this work we derive state-of-the-art model-independent constraints on cosmology from SN Ia by measuring purely kinematical $\left( q,j \right)$ model parameters (where $q$ and $j$ are related to the first and second derivative of the Hubble parameter). For the JLA compilation of SN Ia an agreement within 2$\sigma$ of $\Lambda$CDM expectations is found, where best-fitting kinematical parameters are $q=-0.66 \pm 0.11$ and $j=0.41 ^{+0.32}_{-0.33}$. With $q=-0.73 \pm 0.13$ and $j=0.76 ^{+0.41}_{-0.43}$
the Pantheon sample shows even better agreement with the $\Lambda$CDM expectation of $j=1$ than JLA, hinting at less systematics and/or a higher number of SN Ia alleviating tensions. For the future we predict the precision achievable with SN Ia from the LSST deep survey as $\Delta q \sim 0.05$ and $\Delta j \sim 0.1$, which is systematics-limited and could lead to detect both deviations from $\Lambda$CDM (in $j$) or current expansion rates measured (in $q$). In comparison, for standard cosmological parameters we get $\Delta \Omega_\mathrm{m}=0.01$ and $\Delta w=0.07$ for LSST. 
Given the high number of SN Ia expected for LSST, kinematical parameters in up to 500 sky regions, each with their own individual Hubble diagram, can be constrained. For each region an individual precision at the 10s of percent level is within reach at current systematics-levels, comparable to present-day full-sky surveys. 
This will determine anisotropy in cosmic expansion, or the dark energy dipole, at the 10s of percent level at 10s of degree scales.
\end{abstract}


{
\let\clearpage\relax
\maketitle
}

\section{Introduction}
Deciphering the cause of accelerated cosmic expansion has been a continuous enterprise over the last 20 years, since its first conclusive evidence by means of supernovae Ia (SN Ia) by~\citet{Riess:1998cb} and~\citet{Perlmutter:1998np}. More and more cosmological probes have confirmed this picture since, like measurements of the Cosmic Microwave Background (CMB) by the {\it Planck} satellite~\citep{Ade:2015xua}, as well as galaxy clustering and the abundance of galaxy clusters~\citep{2013MNRAS.432..973R, Mantz14, 2015MNRAS.446.2205M,2016ApJ...832...95D,2018MNRAS.476.4662V,2018arXiv181002322A}. 
Alongside, different scenarios to explain cosmic expansion have been investigated, the most simple being a cosmological constant $\Lambda$ together with Cold Dark Matter (CDM) and assuming General Relativity (GR) to hold, with expansion being both isotropic and homogeneous. A wealth of alternative models have been proposed to explain this accelerated expansion, like scalar-tensor models, for example of the Horndeski class~\citep{1974IJTP...10..363H}. Other models, like Bianchi type I models~\citep{10.2307/1969567}, result in a break-down of the standard assumptions of isotropy in a Friedmann-Lemaitre-Robertson-Walker (FLRW) framework, while trying to accommodate observations.

Different frameworks, like the testing for hemispherical asymmetries or the fitting of dipolar modulations~\citep{Kalus:2012zu,Javanmardi:2015sfa,Ade:2015hxq,Hurier:2016fjp}, 
 strive to detect anisotropies in cosmological data. 
 Works on testing the anisotropy with SN Ia data, as for example by~\citet{Heneka14a,2014IAUS..306...19H} with a Bayesian model-independent approach related to internal robustness of the dataset, or for example~\citet{Cai:2013lja} and~\citet{Sun:2018epo} with standard fitting of cosmological parameters,~\citet{Wang:2014vqa} for cosmographic (kinematical) parameters, and~\citet{Wang:2017ezt} for a Bianchi-I type metric, found no significant evidence of anisotropies in SN Ia data (within 2$\sigma$). 
  Here we aim for a study of kinematical model properties, related to derivatives of the scale factor (see for example~\citet{Frieman:2008sn}), as a model-independent means of constraining acceleration and changes in acceleration of expansion. 

We use both existing SN Ia from the joint light-curve analysis (JLA) by~\citet{Betoule:2014frx} and the Pantheon sample from~\citet{Scolnic:2017caz}, as well as mock realisations of the upcoming Large Synoptic Telescope (LSST) survey of SN Ia, to measure our kinematical model parameters as a global consistency check with $\Lambda$CDM. Furthermore, the upcoming LSST survey of SN Ia~\citep{2009LSST} will measure around 500,000 SN Ia over a large fraction of the sky. This enables us to test individual Hubble diagrams of SN Ia in different directions, splitting the sky up, with a number of SN Ia in each patch comparable to present-day surveys. Here we will investigate how precise we will be able to measure such Hubble diagrams in different regions, where patch-wise parameter deviations would hint at anisotropies in our cosmology. 
Kinematical parameters represent a model-independent set of cosmological parameters, suitable to test global expansion properties and its isotropy. Deviations from simple kinematical $\Lambda$CDM predictions would hint at cosmology beyond, or modifications of gravity like the $f\left(R\right)$ type~\citep{2014PhRvD..90d4016C}. In addition to SN Ia, these kinematical properties will also be testable with redshift drift measurements, where first and second redshift derivatives again are a powerful probe of $\Lambda$CDM cosmology~\citep{2016PhRvD..94d3001M}.

This paper is organised as follows. In section~\ref{sec:kinDE} we describe our kinematical cosmological model and review the framework to derive parameter constraints with apparent magnitudes of SN Ia. In section~\ref{sec:obs} we present constraints on kinematical cosmology derived for present-day data, as well as forecast future constraints attainable with LSST. We continue with a forecast of the precision in kinematical model parameters attainable when measuring Hubble diagrams for different sky regions in section~\ref{sec:LSSTpatch} and finish in section~\ref{sec:out} with our conclusions.

\section{Kinematical cosmological models} \label{sec:kinDE}

Dynamical approaches to constraining cosmology aim at deriving cosmological model parameters, for example the present-day density parameter of dark energy and the dark energy equation of state. In contrast, the kinematical approach relies on in the study of the accelerated background expansion via derivatives of the scale factor $a$ and therefore presents a model-independent alternative to the dynamical approach. It can be based on weaker assumptions, requiring only that gravity is described by some metric theory and that space-time is isotropic and homogeneous. The FLRW metric and the evolution equations for the scale factor $a\left(t\right)$ are still valid~\citep{Frieman:2008sn}.

\subsection{The kinematical approach}
The kinematical parameters up to third order in a Taylor expansion of the scale factor $a\left(t\right)$ are the Hubble parameter $H\left(t\right)$, the deceleration parameter $q\left(t\right)$ and the j-parameter $j\left(t\right)$ that measures the change in acceleration (or deceleration). 
The deceleration parameter, historically defined with a negative sign, measures the cosmic acceleration via
\begin{equation}
q\left(t\right) = \frac{\ddot{a}/a}{\dot{a}^2/a^2} = -1 - \frac{\dot{H}}{H^2} \, ,
\end{equation}
and in terms of the scale factor is
\begin{equation}
q\left( a\right) = -\frac{1}{H} \left( a H\right)'  \, ,
\end{equation}
where the dot denotes derivatives after time $t$ and the prime after scale factor $a$.
Models with present-day q-values $q_0 <0$ currently undergo acceleration. The $j$-parameter, which represents the change in acceleration as the dimensionless third-order time derivative of $a$, is given by
\begin{equation}
j\left(t\right) = -\frac{1}{a H^2} \dddot{a}  \, ,
\end{equation}
and in terms of the scale factor reads
\begin{equation}
j\left( a\right) = -\frac{\left( a^2 H^2\right)''}{2H^2}   \, . \label{eq:ja}
\end{equation}
For a pressure term parametrised via the equation of state that is constant with time, e.g. either matter domination or the domination of a cosmological constant, i.e. in a $\Lambda$CDM scenario, we have $j=1$; for a time evolving pressure term we have $j\neq 1$. The $\Lambda$CDM, or equivalently $j=1$, case presents the zeroth order model around which we are perturbing. The constant $j$ model  captures changes in the accelerated expansion of the Universe at a certain epoch, e.g. for the low redshift Universe with SN Ia. However, for a more realistic treatment a time evolution of $j$ can also be considered.

For convenience, equation~\eqref{eq:ja} can be rewritten as~\citep{Blandford:2004ah,Rapetti:2006fv}
\begin{equation}
a^2 V''\left(a\right) -2j\left( a\right)V\left(a\right) = 0 \label{eq:Vaj} \, ,
\end{equation}
where
\begin{equation}
V\left(a\right) = -\frac{a^2H^2}{2 H_0^2} \, . \label{eq:Va}
\end{equation}
Inserting at present time $a_0=1$ and $H=H_0$, this yields the solution of equation~\eqref{eq:Vaj} with the initial conditions $V\left( 1\right)=-0.5$ and $V'\left( 1\right)=-H'_0/H_0-1 = q_0$. Staying for now with a model that allows for a constant deviation of the $j$-parameter from the $\Lambda$CDM value of $j=1$, equation~\eqref{eq:Vaj} can then be solved analytically to give
\begin{equation}
V\left(a\right) = -\frac{\sqrt{a}}{2} \left[ \left(\frac{p-u}{2p}\right) a^p + \left(\frac{p+u}{2p}\right) a^{-p} \right] \label{eq:Vana} \, ,
\end{equation}
with $p \equiv \left(1/2\right)\sqrt{\left( 1+8j\right)}$ and $u\equiv2\left(q+1/4 \right)$.

Requiring a Big Bang solution (corresponding to the existence of a solution to $V\left(a\right)=0$ in the past as shown in~\citet{Rapetti:2006fv} leads to the exclusion of the following region in the $\left( q,j \right)$ parameter space:
\begin{eqnarray}
&j < q +2q^2  &   q <  -1/4  \, ,  \\ \nonumber
&j < -1/8  & q> -1/4  .
\end{eqnarray}
We will impose these conditions, to exclude regions in parameter space without a Big Bang solution, in our likelihood calculation and parameter estimations in Sections~\ref{sec:obs} and~\ref{sec:LSSTpatch}.

We restrict our analysis to the q-j model described in this section for several reasons. Firstly, equation~(\ref{eq:Va}) presents a consistent and analytical solution without the pitfalls of choosing for example an appropriate expansion~\citep{2019JCAP...01..005M} or time-dependence~\citep{2018arXiv180502854M,2019arXiv190412214P,2019JCAP...05..026G}. Furthermore, one can simply detect and test for possible deviations from $\Lambda$CDM via deviations from $j=1$ (which is also impacted by higher-order corrections), such a detection being a smoking-gun for cosmologies beyond $\Lambda$CDM. Quite importantly, the q-j model is the kinematical equivalent of the standard $w$CDM scenario, which can be translated and compared to each other, which we will do as well in the following. Also with errors significantly larger in the $q$-$j$ parameter space than in the $w-\Omega_\mathrm{m}$ space, no conclusive evidence has been found for higher-order corrections in existing SN Ia data, see for example~\citet{2017EPJC...77..495M} showing that at the moment kinematical two-parameter expansions are preferred by data. While higher order terms might be detectable with LSST, this will crucially depend on the level and exact treatment of systematics.

\subsection{Relating kinematics to dynamics}
Here we relate for later comparison the kinematical $q$- and $j$-parameters to the standard cosmological ones.
We define the standard Hubble function $H$ as $H^2\left( a \right) = \left( \Omega_\mathrm{m,0}a^{-3} + \left( 1-\Omega_\mathrm{m,0}\right) a^{-3\left(1+w\right)}\right)$ for the late Universe, with present-day matter-density $\Omega_\mathrm{m,0}$ and dark energy equation of state $w$.

We start with the q-model, i.e neglecting terms of order j or higher (therefore describing the kinematic evolution as a function of the deceleration parameter $q$ alone with constant acceleration). The effective equation of state $w$ and the kinematic $q$-parameter in this case are connected via
\begin{equation}
w=-\frac{\left( 1-2 q\right)}{3\left( 1-\Omega_\mathrm{m} a^{-3} \left( H_0/H\right)^2\right)} \, ,
\end{equation}
or equivalently
\begin{equation}
q=0.5\left( 1+3w\left( 1-\Omega_\mathrm{m} a^{-3}\right)\left( H_0/H\right)^2\right) \, .
\end{equation}
At the current epoch this leads to
\begin{equation}
q_0=0.5\left( 1+3w\left( 1-\Omega_\mathrm{m,0} \right)\right) \, .  
\label{eq:omtoq}
\end{equation}
We thus can relate the deceleration parameter within a kinematical approach with standard cosmological parameters of the dynamical approach to describe cosmological evolution.

Taking also the change in acceleration with the $j$-parameter into account, within the so-called $q$-$j$-model, one finds for the relation between kinematical and dynamical parameters
\begin{equation}
j=-0.5\left( 1+3w\right)-3q\left( 1+w\right) \, ,
\end{equation}
or, equivalently, from~\citet{Blandford:2004ah},
\begin{equation}
j\left( a\right)=1+\frac{9w\left( 1+w\right)\left( 1-\Omega_\mathrm{m}\right)}{2\left( 1-\Omega_\mathrm{m}\left( 1-a^{3w}\right)\right)} \, .
\label{eq:wtoj}
\end{equation}
We will make use of these relations in the following to compare results in the kinematical model with standard dynamical ones.

\subsection{Constraining the q-j-model with SN Ia}\label{sec:obsSNIa}
To compare with observational data that are sensitive to the background expansion, like SN Ia, inserting $V\left(a\right)$ from equation~\eqref{eq:Vana} into equation~\eqref{eq:Va} gives the evolution of the Hubble parameter as a function of kinematical parameters. The luminosity distance $d_\mathrm{L}$ then reads
\begin{equation}
d_\mathrm{L} = \frac{c}{a H_0} \int_a^1 \frac{\mathrm{d}a}{E\left( a \right)} = \frac{c}{a H_0} \int_a^1 \frac{a\mathrm{d}a}{2\sqrt{V\left( a \right)}} \, ,
\end{equation}
as $E\left( a \right) = H/H_0 = \left(1/a\right)\sqrt{2 V\left( a \right)}$. The luminosity distance is related to the distance modulus $\mu_i$ of a supernova $i$ at redshift $z_i$ with apparent magnitude $m_i$ and absolute magnitude $M$, for a cosmological model with parameter set $\theta_{j}$, via
\begin{equation}
\mu_\mathrm{th,i} = m_\mathrm{th,i} - M  =  5 \log_{10}d_\mathrm{L}\left( z_{i}; \theta_{j} \right) + 25 + K \, ,
\end{equation}
with $K$ being the so-called K-correction that takes into account that  different parts of the source spectrum are observed at different redshifts. The distance modulus is used for cosmological parameter inference, when measured for example with SN Ia (which are assumed to be standard candles of known absolute magnitude). When measuring apparent magnitudes $m_\mathrm{obs,i}$ of SN Ia, the distance modulus $\mu_\mathrm{obs,i}$ at redshift $z_{i}$ is given by
\begin{equation}
\mu_\mathrm{obs,i} = m_\mathrm{obs,i} - M = 5 \log_{10}\hat{d}_\mathrm{L}\left( z_{i} \right)  ,
\label{mod}
\end{equation}
where $\hat{d}_\mathrm{L}$ is the luminosity distance.\footnote{The hat indicates it being in units of $c/H_{0}$.} 

With a sample of observed SN Ia light-curves the distance moduli are fitted as 
\begin{equation}
\mu_\mathrm{obs,i}=m_\mathrm{obs,i}-\left(M_\mathrm{B}  +\alpha x_{i} - \beta c_i +\Delta M+\Delta_\mathrm{B} \right) , 
\end{equation}
with colour and stretch corrections $c_i$ and $x_{i}$, respectively, global best-fitting parameters $\alpha$ and $\beta$ for colour and stretch scaling, as well as absolute B-band magnitude $M_\mathrm{B}$ and the mass step function $\Delta M$ that accounts for correlations of the B-band magnitude with galaxy host mass~\citep{Betoule:2014frx,Jones:2018vbn}. For the Pantheon sample, the factor $\Delta_\mathrm{B}$ was included to account for predicted biases from simulations~\citep{Scolnic:2017caz}.

To obtain parameter constraints on $\left(q,j\right)$ we minimise, as in the standard cosmological framework, the chi-square function marginalised over absolute magnitude $M$, K-correction $K$ and present-day value of the Hubble constant $H_0$, which is given by
\begin{equation}
\chi^2 = S_2 - \frac{S_1^2}{S_0} \, .
\end{equation}
The sums $S_{n}$ are defined as
\begin{equation}
S_{n}=\sum_{i}^{N'}\frac{\delta m_{i}^n}{\sigma_{i}^2} \,,
\label{eq:sums}
\end{equation}
where $\delta m_{i}=\left(m_{{\rm obs,i}}-m_{{\rm th,i}}\right)$ are the magnitude residuals, i.e.~the differences between observed apparent magnitudes and theoretically expected ones, and $\sigma_i$ is the dispersion of distance moduli.
To standardise our analysis for both existing and future SN Ia data instead of assuming different correlation terms, with $\Lambda$CDM being our reference cosmology, we take the common approach~\citep{2012ApJ...746...85S,2014JCAP...03..029K} to add an intrinsic
dispersion $\sigma_\mathrm{int}$. As noted by~\citet{Betoule:2014frx} this procedure assumes he adequacy of the cosmological model, here $\Lambda$CDM, to describe the data. We take $\sigma_\mathrm{int}\sim0.125$ mag for JLA and $\sigma_\mathrm{int}\sim0.126$ mag for Pantheon, in order to obtain a reduced $\chi^2\sim1$ for the best-fit in a $\Lambda$CDM cosmology for the respective datasets.

\section{Observational constraints on kinematics}\label{sec:obs}
\subsection{Results for current datasets: JLA and Pantheon}
When deriving constraints for the JLA compilation of 740 SN Ia in the redshift range $0.01<z<1.3$~\citep{Betoule:2014frx},\footnote{http://supernovae.in2p3.fr/sdss$\_$snls$\_$jla/ReadMe.html} 
as well as for the Pantheon sample~\citep{Scolnic:2017caz},\footnote{https://archive.stsci.edu/prepds/ps1cosmo/index.html}
which at the moment is the largest combined sample of SN Ia consisting of a total of 1048 SN Ia ranging in redshift from $0.01<z<2.3$, we marginalise over the absolute B-magnitude $M_\mathrm{B}$, as well as vary $\alpha$ and $\beta$ in the likelihood calculation. We account for correlations of B-band magnitude with galaxy host mass by fitting the step function parameter $\Delta M $, where $\Delta M$ assumes a fixed value for stellar masses above $10^{10} M_{\odot}$, and zero otherwise. To obtain the errors on cosmological parameters that we are interested in, we marginalise over nuisance parameters $\alpha$, $\beta$ and $\Delta M$.

For the estimation of the diagonal elements in the covariance matrix that correspond to the dispersion in distance moduli, we take the errors of absolute magnitude, colour and stretch as given by the JLA and Pantheon data releases. In addition uncertainties in the flux measurements, intrinsic scatter, as well as scatter due to peculiar velocities are added, as prescribed in the respective data releases.
For example as described in~\citet{Scolnic:2017caz}, the total error in distance measurements for the Pantheon sample takes into account the photometric error, the uncertainty from the mass step correction, distance bias correction, the uncertainty from the peculiar velocity and redshift measurement, as well as the uncertainty from stochastic lensing and intrinsic scatter.
For both samples the SN Ia light-curve parameters were derived with SALT2~\citep{Guy:2007dv}.

Using the formalism described in the previous section we find best-fitting marginalised values and 1$\sigma$ confidence intervals of $q=-0.66 \pm 0.11$ and $j=0.41^{+0.32}_{-0.33}$ for the JLA sample. The constraints are consistent with the $\Lambda$CDM expectation of $j=1$ at the $2\sigma$ level for the marginalised parameter value and agree with accelerated expansion for $q<0$. 
For the Pantheon sample we find $q=-0.73 \pm 0.13$ and $j=0.76 ^{+0.41}_{-0.43}$, in even better agreement with the $\Lambda$CDM expectation of $j=1$ than the JLA sample, hinting at less systematics and/or a higher number of SN Ia helping to improve the q-j parameter estimate.
Interestingly, when fixing the nuisance parameters $\alpha$, $\beta$, and $\Delta_\mathrm{M}$ to their best-fitting values in $\Lambda$CDM, we obtain $q=-0.70 \pm 0.18$ and $j=0.52^{+0.58}_{-0.60}$ and  $q=-0.86 \pm 0.07$ and $j=1.13 \pm 0.26$ for JLA and Pantheon, respectively, with best-fitting values closer to the $\Lambda$CDM expectation and slightly larger confidence intervals.

The corresponding 1-, 2- and 3-$\sigma$ confidence contours for $q$ and $j$ are shown for the JLA and Pantheon sample as black contours in Figure~\ref{fig:jlaLSST}, where the horizontal line indicates the $\Lambda$CDM expectation of $j=1$. For the comparison with future possible constraints by means of the LSST survey of SN Ia see the following section and table~\ref{tab:constraints}.
We also show in Figure~\ref{fig:jlaOmw}, appendix~\ref{app:standard}, the confidence contours derived in a standard $w$CDM scenario for $\Omega_\mathrm{m,0}$ and $w$ for both the JLA compilation and the LSST mock sample. All likelihood calculations are performed numerically on a grid, the chi-square minimisation makes use of the principal axis method of Brent~\citep{Brent:2002}.

\begin{figure}
\begin{center}
\includegraphics[width=0.45\columnwidth]{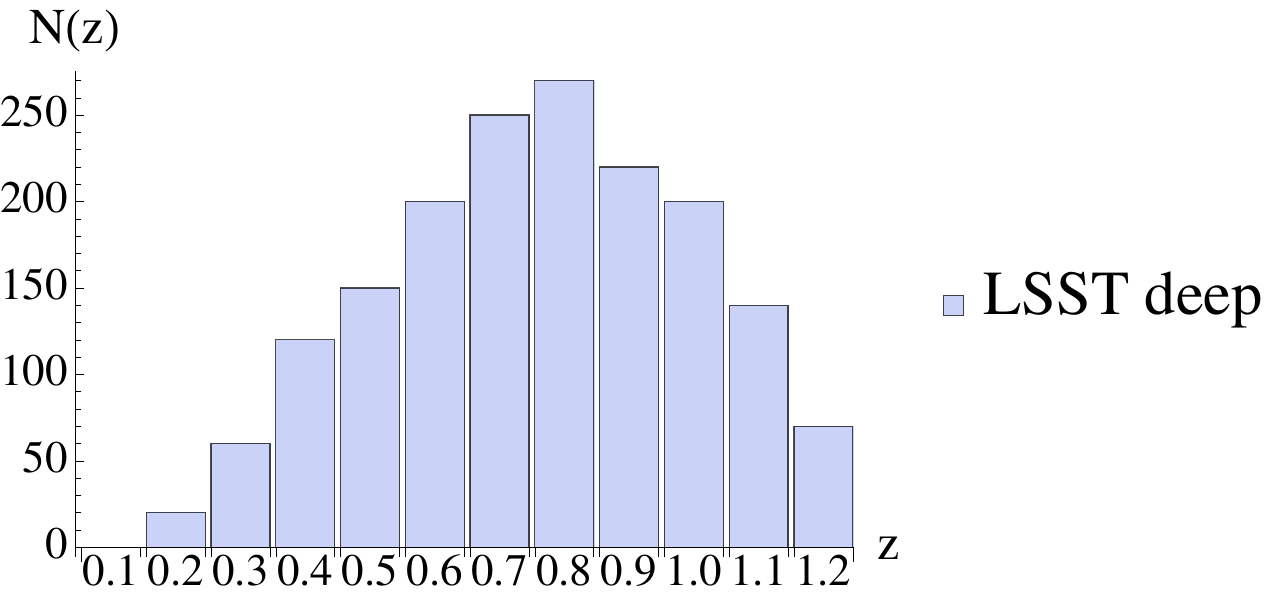} 
\includegraphics[width=0.45\columnwidth]{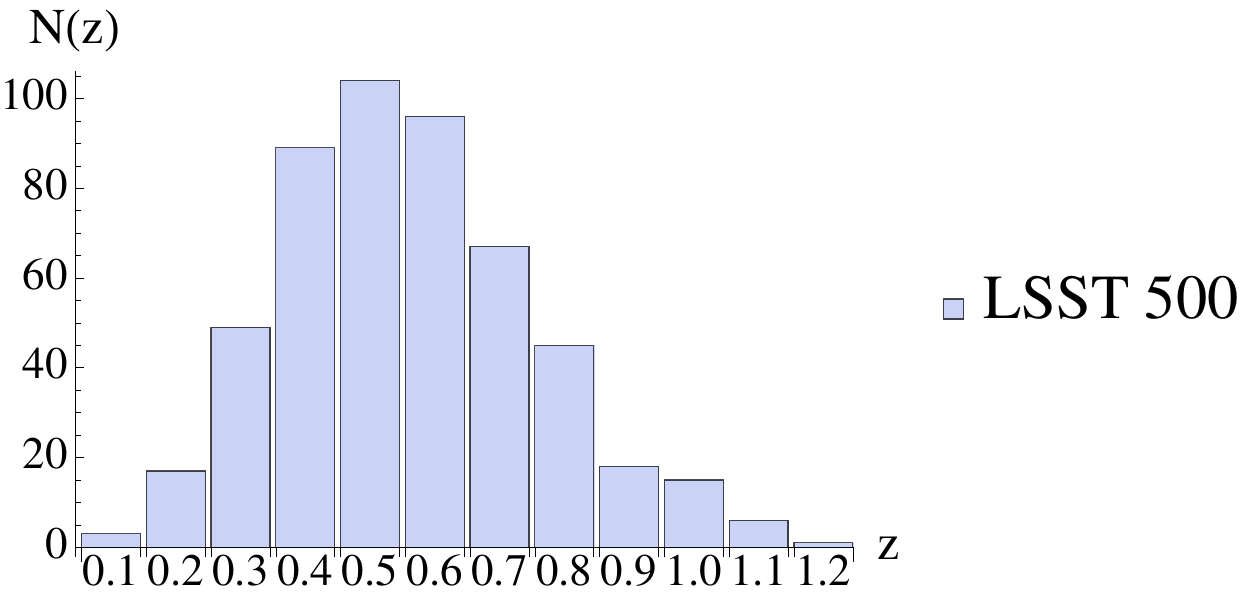} \vspace{0.2cm} \\
\includegraphics[width=0.6\columnwidth]{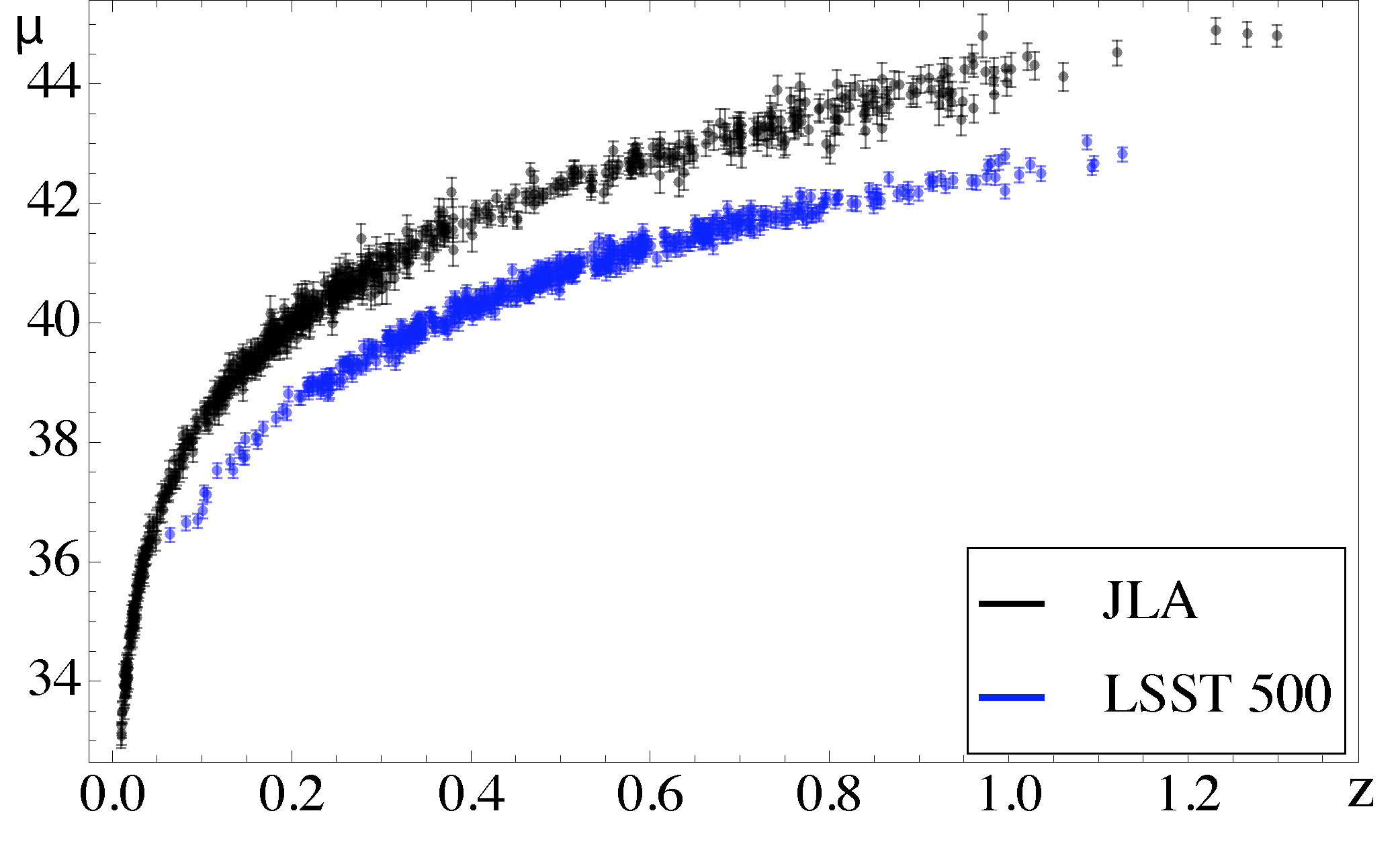}
\caption{Top left: Expected number count binned in redshift for SN Ia detected with the deep LSST survey. Top right: Expected number count binned in redshift for SN Ia detected for one out of 500 sky patches of 40 deg$^2$ for the full LSST survey.
\newline Bottom: Hubble diagrams showing the distance modulus for each SN Ia for the JLA sample (black) and one out of 500 patches of 40 deg$^2$ as part of the full mock LSST SN Ia survey (blue, shifted down for visibility by 1.0$\,$mag); for details on the LSST mock creation see section~\ref{sec:mock}.}
\label{fig:LSSTz}
\end{center}
\end{figure}


\subsection{Future constraints with LSST}\label{sec:LSST}
\subsubsection{Creation of LSST mock SN Ia catalogues}\label{sec:mock}
To investigate constraints of kinematical parameters that will be possible with upcoming SN Ia surveys, we create mock catalogues for the LSST set of SN Ia, both for the full LSST and the LSST deep field.\footnote{https://www.lsst.org/lsst$\_$home.shtml} To do so, we take the predicted redshift distribution for the full LSST and the LSST deep field from~\citet{2009LSST} and calculate the number of SN Ia expected to be observed per year in more than two filters and with a selection cut of signal-to-noise $S/N>15$. For a ten year period of observations this gives the number counts binned in redshift for LSST deep as shown in the top left panel of Figure~\ref{fig:LSSTz}. The top right panel of Figure~\ref{fig:LSSTz} shows the number counts binned in redshift for one out of 500 sky patches for the full LSST survey. It becomes for example obvious how the LSST deep survey will tend to probe more SN Ia at higher redshifts as compared to the full survey. Note as well, that the full LSST survey will produce as many SN Ia measurements for 500 sky regions as do present-day full-sky surveys of SN Ia.

A best-fitting cosmology of $\Omega_\mathrm{m,0}=0.29$ and $w=-1.0$ (the $\Omega_\mathrm{m,0}$ value is derived from JLA for the $\Lambda$CDM expectation with $w=-1.0$ fixed) and kinematical best-fitting parameters of $q=-0.57$ and $j=1.0$ (derived by calculating the corresponding $q$ and $j$ value for our fiducial $\Omega_\mathrm{m,0}=0.29$ and $w=-1.0$ with equations~\eqref{eq:omtoq} and~\eqref{eq:wtoj}) is assumed, as well as a value of the Hubble parameter $h=0.7$ throughout the paper. 
We create mock catalogues by drawing for the chosen fiducial cosmology distance moduli under the expected redshift distribution with a random Gaussian error of $0.05\,$mag, which is the distance modulus dispersion as predicted for the LSST deep field from~\citet{2009LSST}, as well as with an error on distance moduli of $0.12\,$mag for the full LSST field. To underline how exquisite even Hubble diagrams for measurements of 40$\,$deg$^2$ sky regions will be for the full LSST survey, we show in the bottom panel of Figure~\ref{fig:LSSTz} the mock Hubble diagram for one of our 500 sky patches (in blue, with an offset of -1$\,$mag to increase visibility) alongside the JLA Hubble diagram (in black). We will now discuss in the following sections the precision attainable on kinematical model parameters for both the LSST deep field (section~\ref{sec:LSSTdeep}) and for sky patches as part of the full LSST survey in order to test anisotropy of cosmological model parameters in section~\ref{sec:LSSTpatch}.

\subsubsection{q-j constraints from LSST deep}\label{sec:LSSTdeep}
Here we constrain the errors on kinematical $q$- and $j$-parameters attainable with the LSST deep survey, using a mock catalogue of SN Ia distance moduli created as described in the previous section.

\begin{figure}
\begin{center}
\includegraphics[width=0.47\columnwidth]{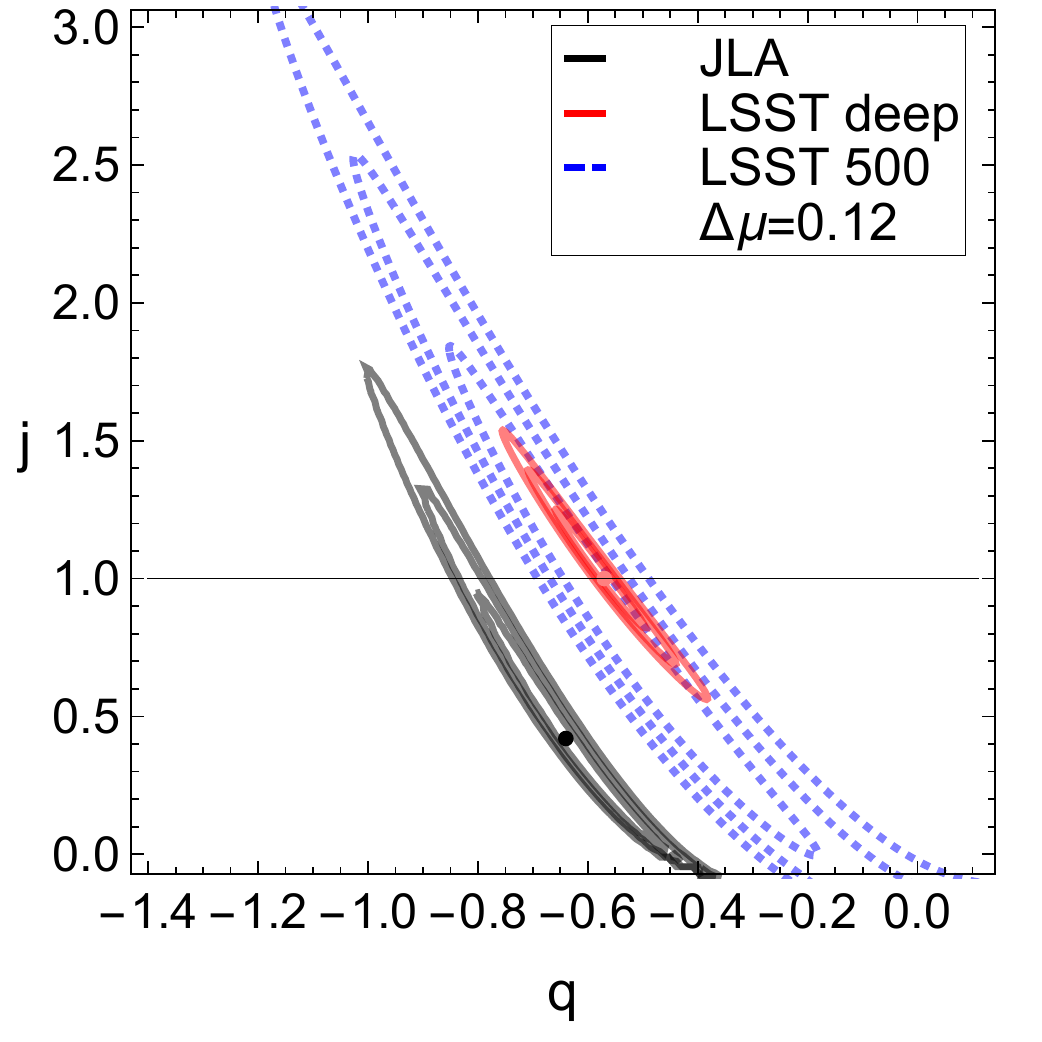}
\includegraphics[width=0.47\columnwidth]{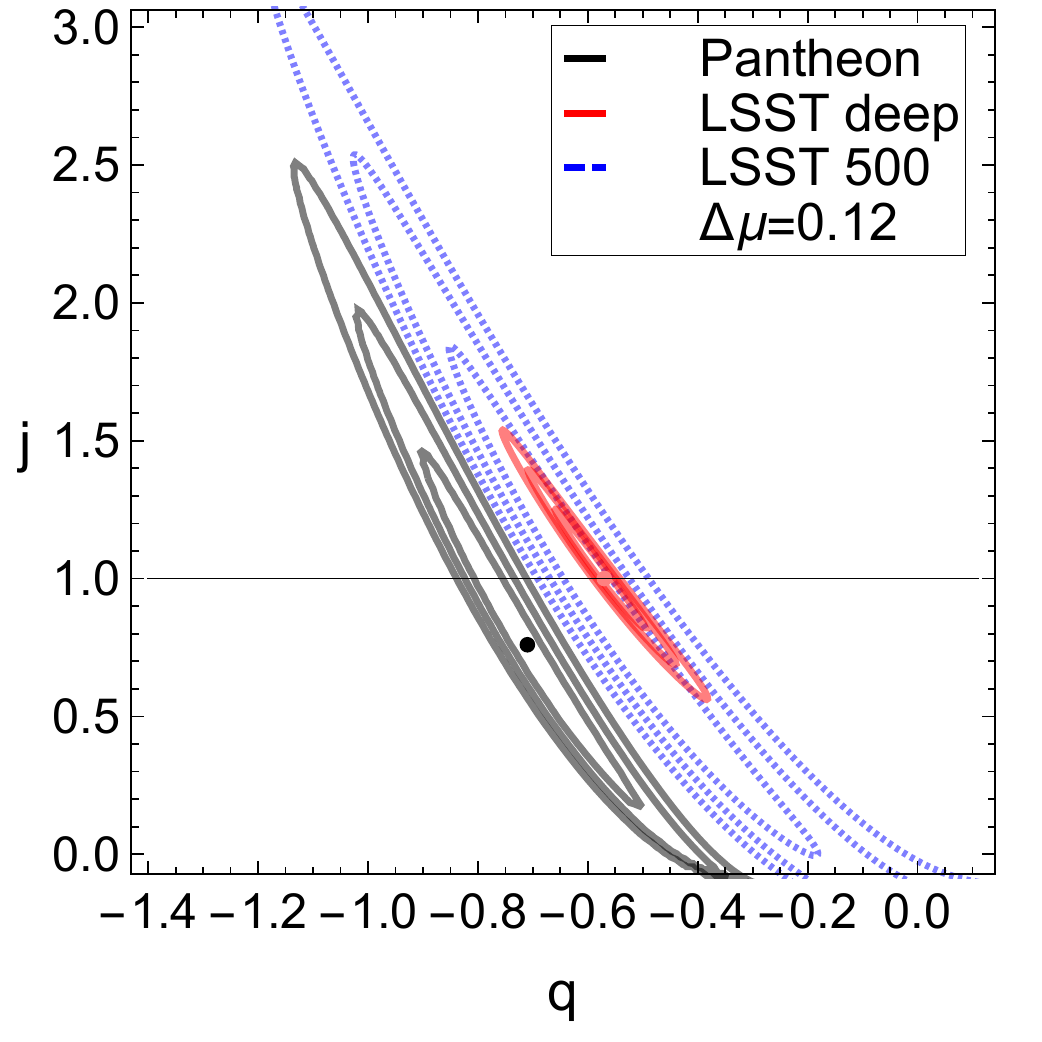}
\includegraphics[width=0.47\columnwidth]{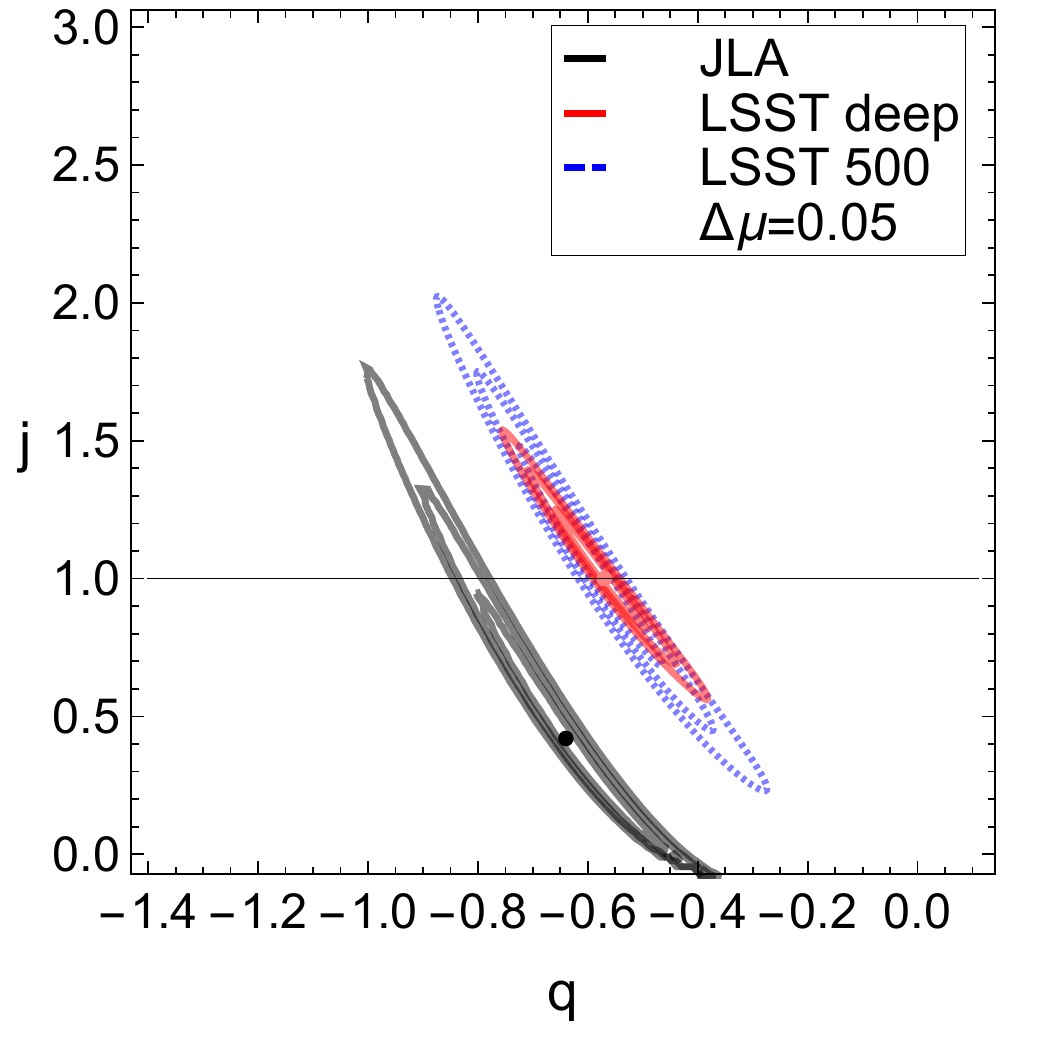} 
\includegraphics[width=0.47\columnwidth]{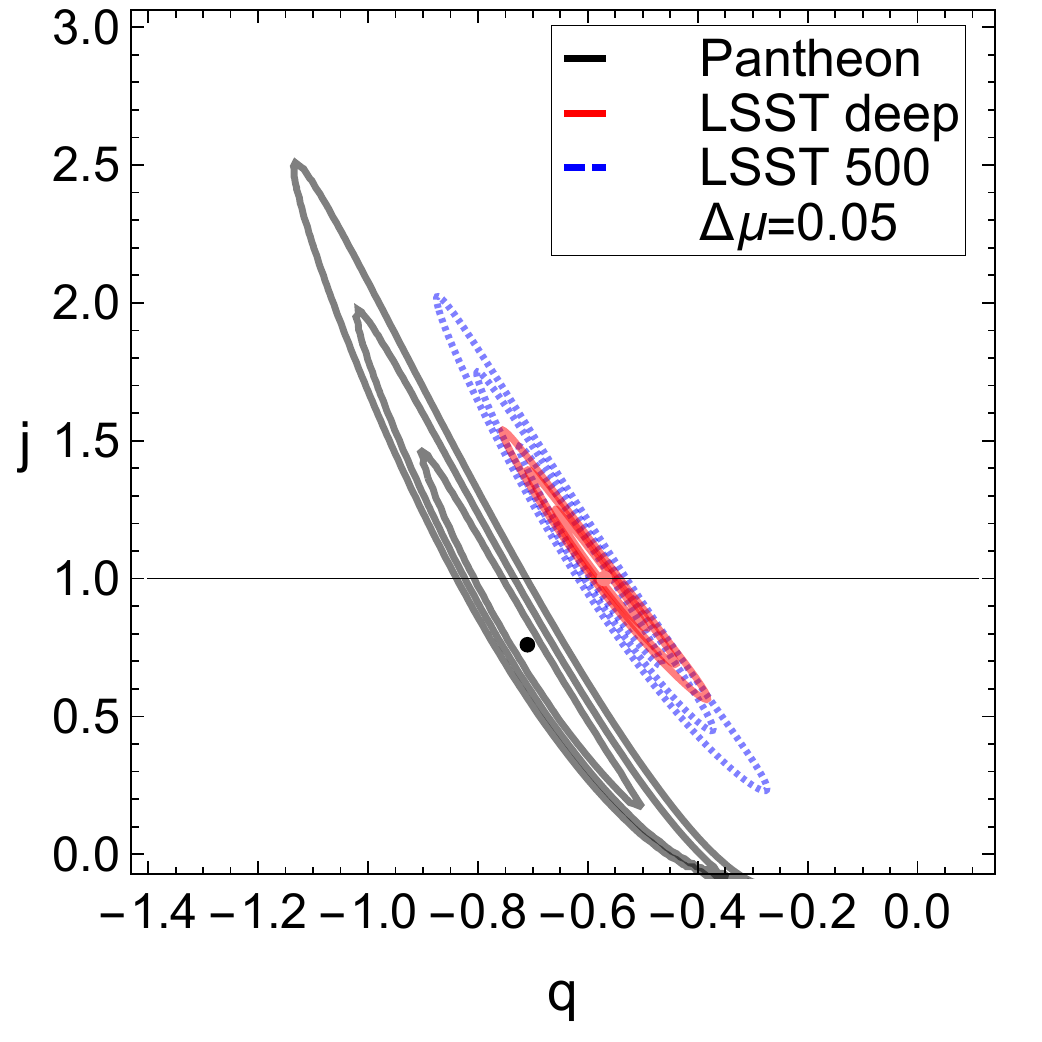}
\caption{Constraints on kinematical model parameters $\left(q,j \right)$ for the JLA (left panels) and Pantheon (right panels) samples of SN Ia (in black) with the best-fit indicated as a black point at $q=-0.66$ and $j=0.41$ for JLA (left) and at $q=-0.73$ and $j=0.76$ for Pantheon (right). Forecasted constraints on kinematical model parameters $\left(q,j \right)$ for one LSST 500 field mock catalogue of SN Ia (dashed blue), assuming an error on the distance indicator of $\Delta\mu =0.12\,$mag (top) and $0.05$mag (bottom), as well as constraints on the LSST deep field with an expected error of $0.05\,$mag (red) are shown alongside. The fiducial model for LSST catalogues $q=-0.57$ and $j=1.0$ is indicated with a red dot, the $\Lambda$CDM prediction of $j=1$ with a black horizontal line. Contours indicate the 68.3, 95.4 and 99.7\% confidence regions.}
\label{fig:jlaLSST}
\end{center}
\end{figure}

For the fiducial model of $q=-0.57$ and $j=1.0$ (the kinematical equivalent of $\Omega_\mathrm{m}=0.29$ and $w=-1.0$), we created the LSST deep mock catalogue of distance moduli and then use this catalogue to constrain the likelihood as described in section~\ref{sec:obsSNIa}.
The corresponding confidence contours in Figure~\ref{fig:jlaLSST} (red for LSST deep) show, $1\sigma$ errors smaller than $\Delta q\approx 0.05$ and $\Delta j\approx 0.1$ are within reach with LSST deep (with an assumed error on distance moduli of $\Delta \mu =0.05$), 
even more for the full survey, which will be systematics-limited. Here we simply assumed an overall level of dispersion motivated by predictions from the LSST Science Book, which we will vary in the following section. Note that especially for LSST systematic effects like catastrophic
photometric errors and the miss-classification of supernovae could become important.

The level of precision reached by LSST opens up ample possibilities, for example testing modifications of GR in different directions of the sky, as we then can divide our supernovae sample into different sky patches, without losing precision. Also possible higher-order extensions of kinematical parameters as well as a time evolution of parameters, when feasible mappable to motivated physical models beyond $\Lambda$CDM, might be testable with LSST if systematics can be pushed down. Testing for time-evolving models beyond LCDM will pose challenges though when for example a strong redshift-dependence in photometric errors will be present.

For marginalised best-fitting values and confidence contours we find $q=-0.58 \pm 0.05$ and $j=1.03 \pm 0.14$ for our mock catalogue, assuming fiducial model parameters $q=-0.57$ and $j=1.0$ as well as dispersion of distance moduli $\Delta \mu = 0.05$.
For estimating the standard cosmological parameters $\left( \Omega_\mathrm{m},w \right)$ from the same mock LSST deep catalogue, we obtain best-fitting values and marginalised 1$\sigma$ errors of $\Omega_\mathrm{m}=0.29 \pm 0.01$ and $w=-1.02 \pm 0.07$, in accordance with a slightly higher precision at the percent level forecasted for the full LSST survey~\citep{2018arXiv180901669T}. Similar to the kinematical parameters, also the standard cosmological parameters can be measured at significantly higher precision than previously with SN Ia, yielding marginalised errors at the percent level on cosmological standard parameters, comparable for example to the precision forecasted for next-generation galaxy cluster studies~\citep{2018arXiv181010553G}, or the Euclid galaxy redshift survey~\citep{2013LRR....16....6A}.

\renewcommand{\arraystretch}{1.5}
\setlength{\tabcolsep}{20pt}
\begin{table}
\centering
\begin{tabular}{| l || c| c| c| c| c| }
\hline \hline
 & $\chi^2_\mathrm{red}$  & $q_\mathrm{BF}$ ($q_\mathrm{fid}$) & $\Delta q$ & $j_\mathrm{BF}$ ($j_\mathrm{fid}$)    & $\Delta j$  \\
\hline 
$\mathrm{JLA}$  & 1.12 &  -0.66 &  $\pm$0.11 &  0.41  &  $^{+0.32}_{-0.33}$  \\ 
$\mathrm{Pantheon}$ & 0.40 & -0.73   &  $\pm$0.13 &  0.76 &  $^{+0.41}_{-0.43}$ \\ 
$\mathrm{LSST deep}$ & 1.07 & (-0.57) & $\pm$ 0.05 & (1.00)  & $\pm$ 0.14  \\  
 $\mathrm{LSST 500 (\Delta \mu}$=0.12) & 1.04 & (-0.57) & $\pm$ 0.30 & (1.00)  & $\pm$ 0.90  \\ 
 $\mathrm{LSST 500 (\Delta \mu}$=0.05) & 1.10 & (-0.57) & $\pm$ 0.13 & (1.00)  & $\pm$ 0.42  \\  
 \hline
     \end{tabular}
        \caption{Summary of best-fitting (fiducial) values for $\left(q,j\right)$-parameters, with marginalized 68.3 per cent confidence intervals $\left(\Delta q,\Delta j\right)$ for SN Ia compilations as indicated.}
\label{tab:constraints}
\end{table}

\section{Dark Energy dipole measurement with LSST}\label{sec:LSSTpatch}

Over its 10 years of operation the LSST will measure an all-sky sample of about 500,000 SNa Ia, which makes it possible to investigate angular dependence in the redshift-distance relation~\citep{2009LSST}. The detection of an angular dependence would point towards a directional dependency of the dark energy equation of state, in turn pointing to physics beyond $\Lambda$CDM. In the past, for example, residuals with respect to the best-fitting Hubble function or hemispherical best-fits have been measured, as done extensively for different samples of SN Ia as mentioned in the introduction, due to the restriction to a low number of SN Ia in each sky patch and their inhomogeneous distribution. With LSST we will be able to constrain a Hubble diagram for a multitude of directions in the sky and therefore test with SN Ia as standard candles the paradigm of isotropic expansion in cosmology. 

In this study we aim to test at what precision kinematical parameters will be measurable in sky patches the size of 40 deg$^2$ predicted to be accessible by LSST~\citep{2009LSST}. This determines the level at which isotropy will be testable via parameter constraint variations patch by patch.
Assuming the SN Ia to be roughly isotropically distributed, we divide the sky into 500 patches of 40 deg$^2$, with 510 SN Ia per patch. The assumption of isotropic distribution of SN Ia and a uniform LSST angular selection function is idealised, but previous analyses (e.g.~\citet{Heneka14a}) have shown that even for SN Ia data as highly non-uniformly distributed as Union2.1, under robust analysis no anisotropy is detected.

We first assume a systematic error of $0.12\,$mag as predicted for the full LSST survey. An error on the kinematical model parameters $q$ and $j$ of $\Delta q = 0.3$ and $\Delta j = 0.9$ per patch is obtained in this configuration, see the corresponding likelihood contour for one single patch in dashed blue in Figure~\ref{fig:jlaLSST} (top panels). 
If we assume an error comparable to the LSST deep survey of $0.05\,$mag to be achievable, constraints are improved with errors at the level of $\Delta q = 0.13$ and $\Delta j = 0.42$ per patch, even outperforming full surveys like JLA. For the corresponding likelihood contours see the blue dashed contours, bottom panels, in Figure~\ref{fig:jlaLSST}. See also the comparison of constraints on $q$ and $j$ parameters in table~\ref{tab:constraints}. The error on parameters $q$ and $j$ per patch is driven by the systematic uncertainty on the distance moduli. This means that for a more accurate testing of isotropy with LSST, systematics would need to be improved on. While we simply assumed and varied an overall level of systematics motivated by earlier LSST predictions, it should be stressed that systematic effects that become crucial and whose impact remains to be investigated in the LSST analysis are for example catastrophic
photometric errors and the miss-classification of supernovae due to the photometry-only nature of the survey.

For comparison, concerning errors on standard dynamical parameters $\Omega_\mathrm{m,0}$ and $w$ for an error of $0.12\,$mag on distance moduli, we derive 1-$\sigma$ errors of $\Delta \Omega_\mathrm{m} = 0.04$ and $\Delta w = 0.12$. This constrains the present-day dark energy dipole at the level of percent to tens of percent. One therefore obtains for standard cosmological parameters, like for kinematic ones, and for each out of 500 sky patches of 40$\,$deg$^2$, constraints that are competitive with present-day constraints from full SN Ia surveys.

\section{Conclusions and Outlook}\label{sec:out}

In this work we have shown that with the upcoming full LSST sample of SN Ia the assumption of isotropy can, not only for standard cosmological parameters, but also kinematical parameters, be tested at unprecedented precision. Besides proving the feasibility of testing isotropy of kinematical parameters with LSST, we also estimated present-day best-fitting values for kinematical parameters for the JLA and Pantheon samples of SN Ia. Our results show agreement with the $\Lambda$CDM expectations of $j=1$ within 1-2$\sigma$. We observe a tendency with a growing number of SN Ia per sample available, together with the inclusion of extra corrections for distance biases, for estimated best-fitting parameters to become more consistent with the $\Lambda$CDM expectation of $j=1$.

To test for constraints on anisotropy achievable with LSST, we divided a LSST mock catalogue of SN Ia in 500 patches. We then measured the corresponding Hubble diagram for each patch to show the ability to detect deviations in kinematical, on top of standard cosmological parameters, at the tens of percent precision, while limited by the error on the distance modulus due to systematics. We here took the approach of assuming overall levels of systematics for the SNIa distance moduli, motivated by values from a more detailed analysis by the LSST collaboration, deferring a detailed analysis of the effect on kinematical cosmological parameters of systematical effects like catastrophic photometric errors to a later study with focus on more realistic LSST mock catalogues.

For the deep LSST sample of SN Ia that is designed to measure light-curves with an $\sim0.05\,$mag error instead of $\sim0.12\,$mag for the full LSST sample, the kinematical parameters are shown to be measurable with a precision of about $\Delta q \sim 0.05$ and $\Delta j \sim 0.1$, on top of an expected $\Delta \Omega_\mathrm{m} \sim 0.01$ and $\Delta w \sim 0.07$ for standard $w$CDM cosmology. This precise, and hopefully accurate, measurement of the kinematics of our Universe, will enable us to get a model-independent handle on possible deviations from our standard assumptions of having a $\Lambda$CDM cosmology in an isotropic and homogeneous expanding universe. Having up to 500 Hubble diagrams distributed over the sky, each one comparable to, or even outperforming, present-day SN Ia surveys, for example stringent constraints on anisotropic models of the Bianchi type can be put in the future. The limit on precision here is set by systematics, demonstrating again the current and future need for tools that statistically select biases in data.

\section*{Acknowledgements}

C.H. acknowledges useful discussions and interaction with David Rapetti at the early stages of this project. This work was supported by the SFB-Transregio TR33 ``The Dark Universe". 







\appendix

\section{Constraints on standard cosmological parameters}\label{app:standard}
\begin{figure}
\begin{center}
\includegraphics[width=0.49\columnwidth]{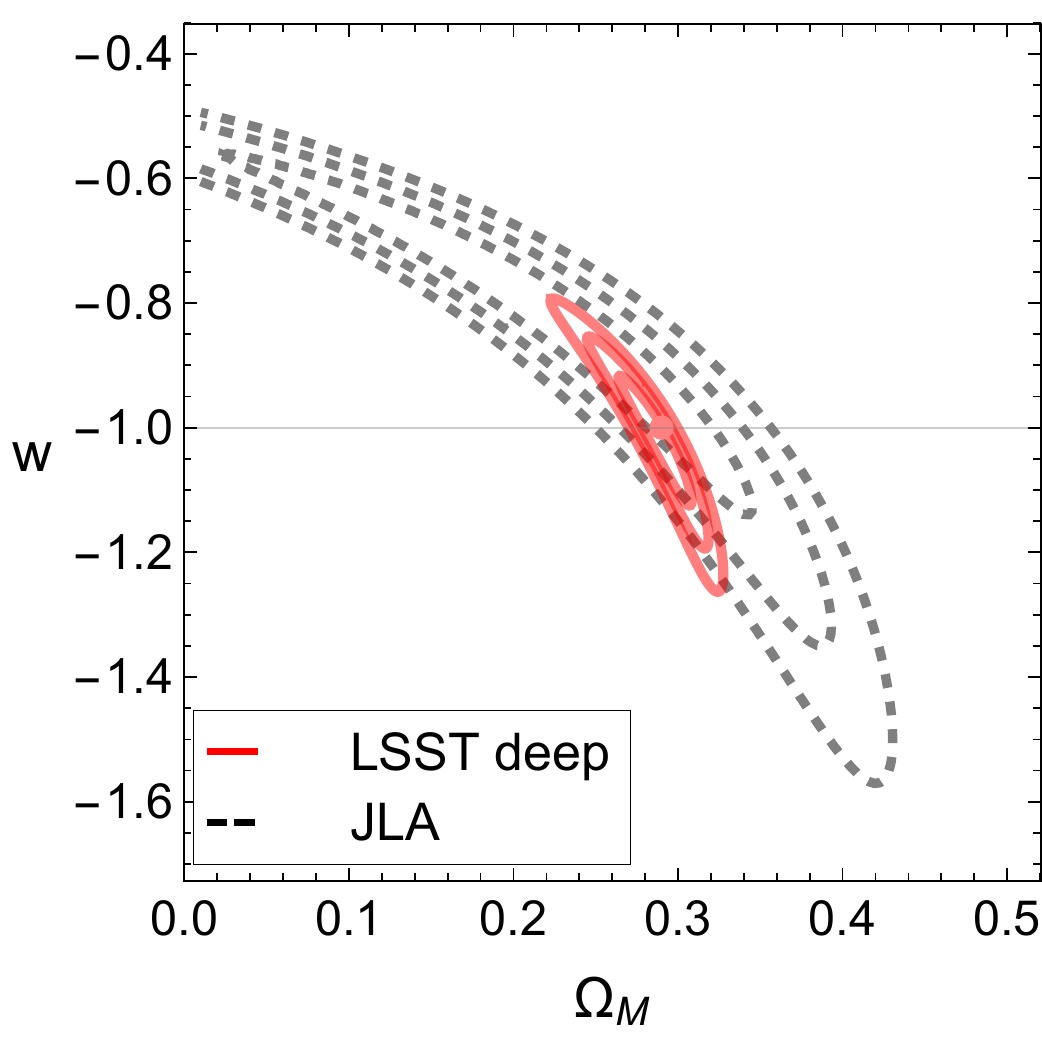}
\caption{Constraints on standard cosmological model parameters $\left(\Omega_\mathrm{m},w \right)$ for the JLA sample (black dashed) and for our LSST deep field mock catalogue (red) of SN Ia. Contours indicate the 68.3, 95.4 and 99.7\% confidence regions. The fiducial LSST model $\Omega_\mathrm{m}=0.29$ and $w=-1.0$ is indicated with a red dot, the $\Lambda$CDM prediction of $w=-1$ with a horizontal line.}
\label{fig:jlaOmw}
\end{center}
\end{figure}

Here we show in Figure~\ref{fig:jlaOmw} for the JLA compilation of SN Ia (black dashed)
constraints on standard cosmological parameters, for comparison with the corresponding contours in the kinematical model in Figure~\ref{fig:jlaLSST}, together with forecasted constraints achievable with LSST deep (in red).  Parameters are $\left(\Omega_\mathrm{m} , w \right)$, with best-fitting marginalised values and 1$\sigma$ confidence intervals of $\Omega_\mathrm{m}=0.264^{+ 0.101}_{- 0.084}$ and $w=-0.81\pm 0.18$ for the JLA sample, 
showing an agreement within 2$\sigma$ with the $\Lambda$CDM expectation. For the LSST deep mock sample, errors of $\Delta \Omega_\mathrm{m} \sim 0.01$ and $\Delta w \sim 0.07$ are within reach around the fiducial of  $\Omega_\mathrm{m} =0.29$ and $w=-1$. Both the kinematical and dynamical approach agree for existing SN Ia  data in their conclusions, agreeing with $\Lambda$CDM expectations at the 1-2$\sigma$-level.

\bibliography{references-q-j}

\end{document}